\documentclass[10pt,conference]{IEEEtran}

\usepackage[utf8]{inputenc}
\usepackage{amsmath,amssymb}
\usepackage{graphicx}
\usepackage{booktabs}
\usepackage{siunitx}
\usepackage{hyperref}
\usepackage{xcolor}
\usepackage{placeins}
\hypersetup{
  colorlinks=true,
  linkcolor=blue,
  citecolor=blue,
  urlcolor=blue
}

\title{End-to-End Throughput Benchmarking of Portable Deterministic CNN-Based Signal Processing Pipelines}

\author{
\IEEEauthorblockN{
Christiaan Boerkamp\thanks{These authors contributed equally to this work.},
Akhil John Thomas\thanks{These authors contributed equally to this work.}
}
\IEEEauthorblockA{
VLV Technology\\
Delft, The Netherlands\\
Email: \texttt{christiaanboerkamp@vlvtechnology.com, ajthomas@vlvtechnology.com}
}
}

\date{January 2026}

\begin{document}
\maketitle

\begin{abstract}
This paper presents a benchmarking methodology for evaluating end-to-end performance of deterministic signal-processing pipelines expressed using CNN-compatible primitives. The benchmark targets phased-array workloads such as ultrasound imaging and evaluates complete RF-to-image pipelines under realistic execution conditions. Performance is reported using sustained input throughput (MB/s), effective frame rate (FPS), and, where available, incremental energy per run and peak memory usage.

Using this methodology, we benchmark a single deterministic, training-free CNN-based signal-processing pipeline executed unmodified across heterogeneous accelerator platforms, including an NVIDIA RTX 5090 GPU and a Google TPU v5e-1. The results demonstrate how different operator formulations (dynamic indexing, fully CNN-expressed, and sparse-matrix-based) impact performance and portability across architectures. This work is motivated by the need for portable, certifiable signal-processing implementations that avoid hardware-specific refactoring while retaining high performance on modern AI accelerators.
\end{abstract}


\section{Introduction}

Modern signal-processing workloads, including ultrasound, radar, sonar, and non-destructive testing (NDT), are increasingly constrained by hardware fragmentation, where each new accelerator platform introduces a distinct programming model and software stack \cite{Wang2019ParaDnn}. In regulated and real-time domains such as medical ultrasound, refactoring, re-validation, and re-certification cycles can be costly and time-consuming.

At the same time, modern accelerators such as high-end GPUs and TPUs provide exceptional performance for convolutional neural networks (CNNs), but often lack efficient native support for classical signal-processing primitives such as dynamic indexing and irregular memory access. This mismatch has been observed in both ultrasound image formation and general accelerator benchmarking studies \cite{Hyun2021CUBDL, Laaber2021Microbenchmarking}.

Recent work has explored the use of deep learning and CNN-based models for ultrasound beamforming and image formation \cite{vanSloun2021DLBeamforming, Goudarzi2022InverseBF, Xiao2024BINN}. While these approaches demonstrate strong reconstruction performance, they typically rely on learned weights and data-driven inference, which can complicate determinism, interpretability, and certification.

Our approach addresses this gap by expressing deterministic, math-defined phased-array signal-processing pipelines entirely using CNN-compatible primitives, including convolutions, pointwise operations, and reductions. These pipelines contain no learned weights and require no training, allowing classical DSP workloads to be executed at accelerator-level performance while preserving determinism, bounded error, and certifiability \cite{Hyun2021CUBDL, Boerkamp2024TINA}.
 The resulting computation graphs can be deployed across heterogeneous hardware targets using a single source-level implementation.

This paper focuses on benchmarking methodology rather than algorithmic derivation. We describe how such pipelines are structured, executed, and measured, and report end-to-end results on contemporary GPU and TPU platforms.

\section{Benchmarking Methodology}

The goal of the benchmark is to measure sustained end-to-end performance of complete signal-processing pipelines rather than kernel-level microbenchmarks. All reported timings include every stage from raw RF input to final image-domain output within a single forward execution.

\subsection{Pipeline Modalities}

The benchmark evaluates three end-to-end imaging modalities:
\begin{itemize}
    \item \textbf{B-mode}: RF $\rightarrow$ IQ $\rightarrow$ beamformed IQ $\rightarrow$ dynamic-range compressed image.
    \item \textbf{Color Doppler}: RF $\rightarrow$ IQ $\rightarrow$ beamformed IQ $\rightarrow$ lag-1 autocorrelation velocity estimation with spatial smoothing.
    \item \textbf{Power Doppler}: RF $\rightarrow$ IQ $\rightarrow$ beamformed IQ $\rightarrow$ power accumulation and log-domain scaling.
\end{itemize}

\subsection{Pipeline Implementation Variants}

For each modality, three implementation variants are benchmarked:
\begin{itemize}
    \item \textbf{Version 1 — Dynamic Indexing}: A reference implementation using explicit indexing or gather-style operations where supported by the backend.
    \item \textbf{Version 2 — Full CNN}: A fully CNN-expressed formulation using only convolutions, pointwise operations, and reductions, designed for maximal accelerator portability.
    \item \textbf{Version 3 — Sparse Matrices}: A formulation that replaces dynamic indexing with structured sparse matrix operations to balance portability and efficiency.
\end{itemize}

This taxonomy enables systematic evaluation of the trade-offs between expressiveness, performance, and hardware portability.

\subsection{Operator Constraints and Determinism}

All pipelines are executed as deterministic forward passes. The forward path is restricted to a controlled operator set comprising element-wise arithmetic, convolutions, pooling or reductions, and simple nonlinearities (e.g., square root and $\operatorname{atan2}$ approximations). No training, stochastic behavior, or data-dependent control flow is present in the benchmarked execution.

All geometry-dependent parameters, lookup tables, and constant kernels are precomputed during module initialization and excluded from timing. The benchmark therefore measures runtime execution of a fixed, fully-initialized pipeline.

\subsection{Input Data and Geometry}

Input RF data are loaded from recorded measurement data and represented as a tensor of dimensions $(n_l, n_c, n_f)$, corresponding to axial samples, receive channels, and temporal frames. A fixed Cartesian image grid and probe geometry are defined prior to execution and reused across all experiments to ensure comparability.

\subsection{Execution Model}

Benchmark execution consists of repeated inference-only forward passes on a fixed input tensor. Gradient computation is disabled. Model parameters and buffers reside on the target device prior to timing. Where applicable, explicit device synchronization is performed to ensure accurate wall-clock measurements. Multiple warm-up iterations are performed before timing to amortize runtime initialization overheads (e.g., caching, compilation, graph setup).

Importantly, the exact same original pipeline codebase is used for both GPU and TPU benchmarks; no algorithmic changes, refactoring, approximations, or backend-specific rewrites are introduced between platforms. Hardware selection is handled exclusively through the execution backend, ensuring that measured differences reflect hardware and runtime characteristics rather than code divergence.

\subsection{Frame Rate Metric}

In addition to throughput, performance is reported in terms of effective frame rate. A frame is defined as a complete RF-to-image forward pass producing a single output image for a given modality. Frame rate is computed as:
\begin{equation}
\text{Frame rate (FPS)} = \frac{1}{T_{\text{avg}}},
\end{equation}
where $T_{\text{avg}}$ is the average forward-pass runtime in seconds. This metric reflects real-time capability under steady-state execution and includes all pipeline stages and synchronization overheads. For B-mode imaging, a single forward pass produces a batch of $N_f=32$ image frames simultaneously, corresponding to the number of temporal frames processed per call. Reported frame rates therefore refer to forward-pass executions per second; the effective image output rate for B-mode is $32\times$ the reported FPS.

\subsection{Throughput Metric}

Throughput is reported as sustained input throughput, defined as the volume of RF input data processed per second during steady-state execution:
\begin{equation}
\text{Throughput (MB/s)} = \frac{B_{\text{in}}}{T_{\text{avg}} \cdot 10^6},
\end{equation}
where $B_{\text{in}}$ denotes the number of input bytes per forward pass and $T_{\text{avg}}$ is the average forward-pass runtime in seconds. This metric captures the end-to-end data ingestion capacity of the complete pipeline and is normalized by input size (rather than intermediate representations) to ensure comparability across variants and hardware backends.

\subsection{Incremental Energy per Run}

For GPU-based experiments, incremental energy consumption per run is reported to quantify the energy cost of executing a single end-to-end pipeline. Energy is estimated by sampling device power during the timed execution window and computing:
\begin{equation}
E_{\text{run}} = \frac{E_{\text{total}}}{N_{\text{runs}}},
\end{equation}
where $E_{\text{total}}$ is the total energy consumed during the timed measurement interval and $N_{\text{runs}}$ is the number of forward passes executed. To isolate workload-related power, an idle power baseline is measured prior to execution and subtracted from active power to form an incremental average power estimate; incremental energy per run is derived from incremental power and runtime.

Energy measurements are reported only for GPU platforms, as board-level power telemetry is not directly accessible for the TPU platform in the current setup.

\subsection{RAM Usage}

Memory usage is reported to characterize runtime footprint. For GPU-based experiments, peak device memory allocation during execution is recorded, including model parameters, constant buffers, intermediate activations, and temporary runtime buffers. Peak memory is measured during steady-state execution, excluding one-time compilation/setup overheads, and reflects the minimum device memory required for continuous pipeline execution.

For TPU experiments, explicit peak memory telemetry is not available in the current setup. TPU memory usage is therefore reported qualitatively based on static model structure and buffer sizes rather than runtime allocation measurements.

\subsection{Hardware Scope}

Benchmarks are conducted on two representative accelerator platforms:
\begin{itemize}
    \item \textbf{GPU}: NVIDIA RTX 5090
    \item \textbf{TPU}: Google TPU v5e-1
\end{itemize}

Dynamic indexing and full-CNN variants are evaluated on both GPU and TPU using identical methodology. The sparse-matrix variant is evaluated on GPU; on TPU it is not reported due to backend operator support limitations.

\begin{table*}[!t]
\centering
\caption{NVIDIA RTX 5090 results (all variants). Input bytes per call: \SI{5.472}{MB}.}
\label{tab:gpu_all}
\scriptsize
\begin{tabular}{@{}l l S[table-format=2.3] S[table-format=4.1] S[table-format=4.1] S[table-format=1.3] S[table-format=1.3]@{}}
\toprule
\textbf{Pipeline} & \textbf{Variant} &
{$T_{\text{avg}}$ (ms)} & {FPS} & {MB/s} & {J/run} & {Peak Mem (GB)} \\
\midrule
RF2IQ\_DAS\_DOPPLER      & Dynamic indexing & 0.787  & 1270.6 & 6953.87 & 0.047 & 0.337 \\
RF2IQ\_DAS\_POWERDOPPLER & Dynamic indexing & 0.754  & 1326.2  & 7256.36 & 0.043 & 0.669 \\
RF2IQ\_DAS\_BMODE        & Dynamic indexing & 11.835 & 84.5   & 462.38  & 3.108 & 1.001 \\
\midrule
RF2IQ\_DAS\_DOPPLER      & Full CNN         & 8.852  & 113.0  & 618.20  & 2.162 & 0.359 \\
RF2IQ\_DAS\_POWERDOPPLER & Full CNN         & 8.949  & 111.7  & 611.49  & 2.152 & 2.391 \\
RF2IQ\_DAS\_BMODE        & Full CNN         & 19.287 & 51.8   & 283.72  & 5.779 & 2.745 \\
\midrule
RF2IQ\_DAS\_DOPPLER      & Sparse matrices  & 19.213 & 52.0   & 284.83  & 4.857 & 0.171 \\
RF2IQ\_DAS\_POWERDOPPLER & Sparse matrices  & 19.215 & 52.0   & 284.79  & 4.864 & 5.950 \\
RF2IQ\_DAS\_BMODE        & Sparse matrices  & 29.574 & 33.8   & 185.04  & 8.321 & 6.116 \\
\bottomrule
\end{tabular}
\end{table*}

\subsection{TPU Benchmarking Methodology}

To evaluate accelerator portability beyond GPU execution, the dynamic indexing and full-CNN variants were benchmarked on a Google TPU v5e-1 using the same original codebase as used for GPU execution. No algorithmic modifications or TPU-specific rewrites were introduced; the same pipeline definitions, operator structure, and execution logic were preserved across platforms.

All TPU experiments use fixed input shapes and a static execution graph. Multiple warm-up iterations are performed to amortize compilation and graph initialization overhead before timing begins. Timing measurements are taken over repeated forward executions, and average runtime is reported. Power and incremental energy are not reported for TPU due to limited board-level power telemetry access in the current setup.

\section{Results}

Results are reported per imaging modality, per implementation variant, and per hardware platform (NVIDIA RTX 5090 and TPU v5e-1). All results are obtained using the same original pipeline implementation and identical benchmarking methodology. Sustained input throughput and frame rate are measured for both platforms, while incremental energy per run and peak memory usage are reported where hardware telemetry is available (GPU).

\subsection{GPU Results (All Variants)}

Table~\ref{tab:gpu_all} reports NVIDIA RTX 5090 results for all three pipeline implementation variants across all evaluated modalities. All measurements use a fixed input size of \SI{5.472}{MB} per forward pass and are obtained using the same original pipeline codebase as used for TPU benchmarking, without algorithmic modification or backend-specific rewrites.

For the dynamic-indexing variant, the GPU achieves very high sustained throughput and frame rates, exceeding \SI{6.9}{GB/s} and \SI{1200}{FPS} for Doppler workloads. These results reflect the strong support for irregular memory access and gather-style operations on modern GPU architectures. However, this variant also exhibits higher sensitivity to memory footprint and intermediate activation materialization, particularly for B-mode.

The fully CNN-expressed variant demonstrates reduced peak throughput relative to dynamic indexing on GPU, but achieves substantially lower memory access irregularity and improved portability. This variant provides a balanced trade-off between performance, determinism, and accelerator compatibility, and serves as a common representation across GPU and TPU platforms.

The sparse-matrix variant exhibits competitive throughput for Doppler workloads but incurs increased peak memory usage, particularly for Power Doppler and B-mode. This reflects the cost of materializing structured sparse representations and intermediate buffers on GPU hardware.

\subsection{TPU Results}

Table~\ref{tab:tpu_all} reports TPU v5e-1 results for the dynamic-indexing and full-CNN variants. All measurements use a fixed input size of \SI{5.472}{MB} per forward pass and are obtained using the same original pipeline implementation as used for GPU benchmarking, without algorithmic modification or TPU-specific rewrites.

For the dynamic-indexing variant, sustained throughput is approximately \SI{30}{MB/s} across all modalities, corresponding to forward-pass rates of approximately \SIrange{5.4}{5.5}{FPS}. These results reflect the limited efficiency of dynamic indexing and irregular memory access on the TPU execution backend.

In contrast, the fully CNN-expressed variant achieves substantially higher performance on TPU, with throughput exceeding \SI{500}{MB/s} and forward-pass rates of up to \SI{104}{FPS}. This demonstrates that expressing classical signal-processing pipelines exclusively using CNN-compatible primitives enables effective utilization of TPU hardware without compromising determinism or portability.

The sparse-matrix variant is not reported for TPU. Structured sparse operators are not fully supported by the current TPU execution backend (\texttt{xm.xla}), preventing faithful execution of the sparse formulation using the same codebase.

\begin{table*}[!t]
\centering
\caption{TPU v5e-1 results for all pipeline variants. Input bytes per call: \SI{5.472}{MB}.}
\label{tab:tpu_all}
\begin{tabular}{@{}l l S[table-format=2.3] S[table-format=4.1] S[table-format=4.1]@{}}
\toprule
\textbf{Pipeline} & \textbf{Variant} &
{$T_{\text{avg}}$ (s)} & {FPS} & {MB/s} \\
\midrule
RF2IQ\_DAS\_DOPPLER      & Dynamic indexing & 0.181 & 5.53  & 30.27 \\
RF2IQ\_DAS\_POWERDOPPLER & Dynamic indexing & 0.182 & 5.48  & 30.00 \\
RF2IQ\_DAS\_BMODE        & Dynamic indexing & 0.184 & 5.43  & 29.70 \\
\midrule
RF2IQ\_DAS\_DOPPLER      & Full CNN         & 0.01 & 96.4  & 527.72 \\
RF2IQ\_DAS\_POWERDOPPLER & Full CNN         & 0.009 & 103.9 & 568.39 \\
RF2IQ\_DAS\_BMODE        & Full CNN         & 0.012 & 83.3  & 455.72 \\
\bottomrule
\end{tabular}
\end{table*}

\section{Discussion}

End-to-end benchmarking captures performance effects that are often invisible to kernel microbenchmarks, including memory traffic between stages, intermediate tensor materialization, and synchronization overhead. Reporting throughput (MB/s), frame rate (FPS), and (where available) incremental energy per run and peak memory provides a deployment-oriented view of performance for real-time and embedded systems.

Comparing the dynamic-indexing baseline across platforms, the NVIDIA RTX 5090 achieves substantially higher throughput and frame rate than the TPU v5e-1 using the same original pipeline implementation. This performance gap reflects differences in memory access patterns, dynamic indexing efficiency, and runtime maturity rather than algorithmic differences, and motivates further analysis of the fully CNN and sparse-matrix formulations on TPU.

Kernel-level microbenchmarks are intentionally avoided, as they fail to capture synchronisation, memory traffic, and intermediate tensor materialisation effects present in deployed pipelines, e.g., via PyTorch's compile functionality.

To contextualize the throughput of our deterministic CNN-based pipelines, we compare sustained input throughput (MB/s or GB/s) with results reported in prior deterministic GPU-based implementations. Throughput is either taken directly from the literature or computed from reported frame rates and input sizes.

Our Doppler pipeline using dynamic indexing achieves a sustained throughput of \SI{7.2}{GB/s} on a single RTX~5090 GPU, matching or exceeding the highest reported values in prior deterministic works. On TPU, the fully CNN-expressed variant reaches \SI{530}{MB/s}, demonstrating strong performance portability without requiring hardware-specific rewrites.

Yiu~\emph{et al.}~\cite{Yiu2018GPUUltrasound} report frame rates of 3,000–4,700~FPS for 2D plane-wave imaging using dual GTX~480 GPUs, corresponding to a throughput of approximately 1–2~GB/s. Rossi~\emph{et al.}~\cite{Rossi2023UltrafastDoppler} achieve 2,500–3,500~FPS for vector Doppler on a Jetson Xavier, reaching an estimated 7–8~GB/s, although throughput was PCIe-limited. Liu~\emph{et al.}~\cite{Liu2023Ultrasound3DRCA} report a 3D row–column beamformer running at 1,394~volumes/s on an RTX~4090, with compressed input throughput of 2.3~GB/s; raw data throughput without compression would exceed 350~GB/s.

In contrast to prior implementations which often involve hardware-specific tuning or compression strategies, our work executes an unmodified deterministic CNN-based pipeline across both GPU and TPU platforms. This demonstrates that deterministic CNN-compatible formulations can achieve state-of-the-art throughput while maintaining cross-platform portability and certifiability.

\begin{table*}[!t]
\centering
\caption{Reported or computed throughput of prior deterministic GPU/TPU implementations.}
\label{tab:comparison_prior}
\begin{tabular}{@{}l l@{}}
\toprule
\textbf{Source / Pipeline} & \textbf{Throughput} \\
\midrule
This work (RTX 5090, Doppler, Dyn. Indexing) & \SI{7.2}{\giga\byte\per\second} \\
This work (TPU v5e-1, Doppler, Full CNN)     & \SI{0.53}{\giga\byte\per\second} \\
Yiu et al.~(2018)\cite{Yiu2018GPUUltrasound} & \SIrange{1}{2}{\giga\byte\per\second} \\
Rossi et al.~(2023)\cite{Rossi2023UltrafastDoppler} & \SIrange{7}{8}{\giga\byte\per\second} \\
Liu et al.~(2023)\cite{Liu2023Ultrasound3DRCA} & \SI{2.3}{\giga\byte\per\second} (compressed) \\
\bottomrule
\end{tabular}
\end{table*}

\section{Limitations}

This work focuses on throughput and runtime benchmarking and does not directly evaluate image quality or clinical performance. While the benchmarking methodology is fully specified, the complete implementation code cannot be publicly released due to proprietary constraints; the emphasis is therefore on transparent methodology and comparative performance trends rather than exact numerical replication.

Additionally, some telemetry (e.g., board-level power and peak memory allocation) is not directly accessible on TPU in the current setup; TPU reporting is therefore limited to runtime-derived metrics such as FPS and MB/s.

\section{Conclusion}

This paper presented a reproducible benchmarking methodology for evaluating end-to-end performance of deterministic signal-processing pipelines expressed using CNN-compatible operations. By reporting sustained input throughput, effective frame rate, and (where available) incremental energy per run and peak memory usage, the benchmark captures deployment-relevant performance characteristics that are not visible through kernel-level microbenchmarks.

Using this methodology, we evaluated complete RF-to-image ultrasound pipelines across multiple implementation variants on two fundamentally different accelerator platforms: an NVIDIA RTX 5090 GPU and a Google TPU v5e-1. Crucially, the same original deterministic CNN-based pipeline codebase was executed unmodified on both platforms. No algorithmic changes, backend-specific rewrites, or approximations were introduced between GPU and TPU experiments; differences in performance therefore reflect hardware architecture and runtime behavior rather than code divergence.

The results demonstrate that fully CNN-expressed formulations enable strong performance portability across heterogeneous accelerators. While dynamic indexing achieves very high performance on GPU hardware, it performs poorly on TPU. In contrast, the fully CNN-based formulation achieves high throughput on both platforms, exceeding \SI{500}{MB/s} on TPU while maintaining deterministic behavior and a static execution graph. These results highlight the value of expressing classical signal-processing workloads using accelerator-native CNN primitives when portability and certifiability are primary design constraints.

Overall, this work shows that deterministic, training-free CNN-based signal-processing pipelines can be benchmarked fairly and executed efficiently across heterogeneous AI accelerators using a single source-level implementation. The presented methodology provides a foundation for evaluating performance portability of classical DSP workloads on current and future accelerator architectures.

\section{Future Work}

The benchmarking results presented in this work motivate several directions for future investigation.

First, a follow-on paper will provide a detailed methodological description of how classically non-CNN signal-processing operations can be expressed deterministically using CNN-compatible primitives. This includes control-flow constructs (e.g., conditional selection), transcendental functions (e.g., logarithms and $\operatorname{atan2}$), and nonlinear magnitude and phase operations. Rather than relying on learned approximations, these operators are implemented using fixed, math-defined compositions of convolutions, pointwise operations, and reductions, preserving determinism and bounded numerical behavior.

Second, we plan to describe in detail how these deterministic CNN-expressed operators are integrated into complete ultrasound processing pipelines. This includes RF-to-IQ demodulation, delay-and-sum beamforming, B-mode dynamic range compression, IQ-domain Doppler velocity estimation, and power Doppler accumulation. The focus will be on demonstrating how traditionally irregular or data-dependent operations can be reformulated into static execution graphs suitable for accelerator deployment, while maintaining algorithmic equivalence to classical signal-processing formulations.

Third, future work will focus on systematic optimization of these CNN-based implementations for both performance and resource efficiency. This includes operator fusion, memory layout optimization, structured sparsity, and backend-specific lowering strategies, while preserving a single source-level implementation across hardware platforms.

Fourth, a dedicated follow-on study will evaluate numerical accuracy and output fidelity of the proposed deterministic CNN-based pipelines. This will include quantitative error analysis against reference signal-processing implementations, sensitivity studies with respect to fixed-point and reduced-precision arithmetic, and modality-specific image quality metrics. Together, these efforts aim to complement the performance benchmarking presented here with a rigorous assessment of correctness and accuracy.

Fifth, we plan to extend benchmarking to additional hardware platforms beyond traditional GPUs and TPUs, including neuromorphic chips, in-memory compute architectures, and other emerging CNN accelerators. Developers of novel hardware platforms interested in evaluating their accelerator on real-time, deterministic DSP pipelines are encouraged to contact the authors for potential collaboration and testing.

Lastly, we are developing a CNN-compatible signal compressor module designed to reduce intermediate tensor volume, enabling execution of large-array ultrasound pipelines on memory-constrained or embedded accelerators. This will facilitate broader deployment of high-fidelity deterministic pipelines in portable and resource-limited devices.

\section*{Reproducibility Statement}

Although the full source code cannot be publicly released, all benchmarks are derived from a single original codebase that is executed unmodified across GPU and TPU platforms. All experiments follow a fixed and clearly defined execution procedure, input specification, and performance metrics (MB/s and FPS; and for GPU, incremental energy per run and peak memory). This enables independent reimplementation and fair comparison of alternative approaches using the same methodology.
\bibliographystyle{IEEEtran}
\bibliography{References}
\end{document}